\documentclass{elsart}
\usepackage{hyperref}

\newcommand{\hepex}[1]{\mbox{\href{http://xxx.lanl.gov/abs/hep-ex/#1}{(hep-ex/#1)}}}
\newcommand{\hepph}[1]{\mbox{\href{http://xxx.lanl.gov/abs/hep-ph/#1}{(hep-ph/#1)}}}
\newcommand{\astroph}[1]{\mbox{\href{http://xxx.lanl.gov/abs/astro-ph/#1}{(astro-ph/#1)}}}
\def\url#1{\mbox{\href{#1}{\sf #1}}}
\def\urll#1#2{\mbox{\href{#1}{\sf #2}}}
\def\urlp#1#2{\mbox{\href{#1}{#2}}}

\def\pr#1#2#3{\frenchspacing{\it Phys. Rev. D}{\bf #1} (19#3) #2}
\def\prstab#1#2#3{\frenchspacing{\it Phys. Rev. ST Accel. Beams\ }{\bf #1} (19#3) #2}
\def\prl#1#2#3{\frenchspacing{\it Phys. Rev. Lett.\ }{\bf #1} (19#3) #2}
\def\pl#1#2#3{\frenchspacing{\it Phys. Lett.\ }{\bf #1} (19#3) #2}
\def\prc#1#2#3{\frenchspacing{\it Phys. Rev. C}{\bf #1} (19#3) #2}
\def\np#1#2#3{\frenchspacing{\it Nucl. Phys. }{\bf #1} (19#3) #2}
\def\ib#1#2#3{\frenchspacing{\it ibid. }{\bf #1} (19#3) #2}
\def\jhep#1#2#3{\frenchspacing{\it J. High Energy Phys.\ }{\bf #1} (19#3) #2}
\def\npbps#1#2#3{{\em Nucl. Phys. B (Proc. Supp.)\/} {\bf #1} (19#3) #2}
\def\half{{\scriptstyle \frac{1}{2}}}     
\def\two3{{\scriptstyle \frac{2}{3}}}     

\newcommand{\lag}{\ensuremath{\mathcal{L}}}

\newcommand{\ewgg}{\ensuremath{SU(2)_L \otimes U(1)_Y}}

\newcommand{\ev}{\ensuremath{\hbox{ eV}}}
\newcommand{\evcc}{\ensuremath{\hbox{ eV}\!/\!c^2}}
\newcommand{\mevcc}{\ensuremath{\hbox{ MeV}\!/\!c^2}}

\newcommand{\gev}{\ensuremath{\hbox{ GeV}}}
\newcommand{\cm}{\ensuremath{\hbox{ cm}}}
\def\ltap{\mathop{\raisebox{-.4ex}{\rlap{$\sim$}} \raisebox{.4ex}{$<$}}}

\def\slashii#1{\setbox0=\hbox{$#1$}             
   \dimen0=\wd0                                 
   \setbox1=\hbox{\sl/} \dimen1=\wd1            
   \ifdim\dimen0>\dimen1                        
      \rlap{\hbox to \dimen0{\hfil\sl/\hfil}}   
      #1                                        
   \else                                        
      \rlap{\hbox to \dimen1{\hfil$#1$\hfil}}   
      \hbox{\sl/}                               
   \fi}                                         %
\def\slashiii#1{\setbox0=\hbox{$#1$}#1\hskip-\wd0\hbox to\wd0{\hss\sl/\/\hss}}
\journal{Nuclear Instruments and Methods A}
\runauthor{Chris Quigg}
\runtitle{Questions of Identity}

\begin{document}


\begin{frontmatter}
\begin{flushright}
\phantom{rufus}\vspace{-48pt}
FERMILAB--CONF--99/233--T
\end{flushright}


\title{Questions of Identity}


\author{Chris Quigg\thanksref{DOE}}
\thanks[DOE]{E-mail address: \textsf{quigg@fnal.gov}.}
\address{Fermi National Accelerator Laboratory \\ P.O. Box 500, 
Batavia, Illinois 60510 USA}


\begin{abstract}
As an introduction to $\nu$Fact '99, the ICFA/ECFA Workshop on 
Neutrino Factories Based on Muon Storage Rings, I place the issues of 
neutrino properties and neutrino oscillations in the broader context 
of fermion flavor.
\end{abstract}




\end{frontmatter}


\section{Introduction}
Our colleagues working to assess the feasibility of very-high-energy 
muon colliders \cite{mucoll} have given us the courage to think that 
it may be possible, not too many years in the future, to accumulate 
$10^{20\mathrm{ - }21}$ (or even $10^{22}$) muons per year.  It is 
very exciting to think of the possibilities that millimoles of muons 
would raise for studies in fundamental physics 
\cite{fmc,cqfmc,autres}, and indeed that is why we have come together 
today in \urlp{http://lyoinfo.in2p3.fr/nufact99/}{Lyon.}

From the perspective of a muon collider, the 2.2-$\mu$s lifetime of 
the muon presents a formidable challenge.  But if the challenge of 
producing, capturing, storing, and replenishing many unstable muons 
can be met, the decays

\begin{equation}
	\mu^{-}  \rightarrow  e^{-}\nu_{\mu}\bar{\nu}_{e}\; , \qquad 
	\mu^{+}  \rightarrow  e^{+}\bar{\nu}_{\mu}\nu_{e}
	\label{mumpdk}
\end{equation}
offer delicious possibilities for the study of neutrino interactions 
and neutrino properties \cite{geer,abp,bgw,suite}.  In a \textit{Neutrino 
Factory}, the composition and spectra of intense neutrino beams will 
be determined by the charge, momentum, and polarization of the stored 
muons.  At the energies best suited for the study of neutrino 
oscillations---tens of GeV, by our current estimates---the muon 
storage ring is compact.  We could build it at one laboratory, pitched 
at a deep angle, to illuminate a laboratory on the other side of the 
globe with a neutrino beam whose properties we can control with great 
precision.  By choosing the right combination of energy and 
destination, we can tune future neutrino-oscillation experiments to 
the physics questions we will need to answer, by specifying the ratio 
of path length to neutrino energy and determining the amount of matter 
the neutrinos traverse.  Although we can use each muon decay only 
once, and we will not be able to select many destinations, we may be 
able to illuminate two or three well-chosen sites from a 
muon-storage-ring neutrino source.  That possibility---added to the 
ability to vary the muon charge, polarization, and energy---may give 
us just the degree of experimental control it will take to resolve the 
outstanding questions about neutrino oscillations.

\section{Some Issues for the Workshop}
As we begin this workshop, it seems to me that we should keep in mind 
four essential questions:
\begin{verse}
	Is a Neutrino Factory feasible? \\[6pt]

	At what cost? \\[6pt]

	How soon? \\[6pt]

	What R\&D must we do to learn whether we can make the neutrino 
	factory a reality?
\end{verse}
The answers to these questions will be influenced by what we want the 
neutrino factory to be.  To decide that, we need to consider another 
set of questions:
\begin{verse}
	What do we want to know about neutrino masses and mixings now \ldots 
	in five years \ldots in ten years? \\[6pt]
	
	Is a neutrino factory the best way---or the only way---to provide 
	this information? \\[6pt]

	What (range of) beam parameters should the neutrino factory 
	offer? \\[6pt]

What detectors are needed to carry out the physics program of a 
neutrino factory?  It seems that distant detectors must weigh several 
kilotonnes and ideally should identify electrons, muons, and taus---and 
measure their charges.  Are all these characteristics essential?  How 
should the prospect of a neutrino factory influence the detectors we 
build now?
\end{verse}

We need to consider the scientific issues with an eye to both the 
intrinsic interest in neutrino properties and interactions and also 
the evolving place of neutrino physics within contemporary particle 
physics.  I find it useful to organize the goals of particle physics 
at the millennium in terms of three broad themes.  The first theme is 
\textit{symmetry,} with the attendant ideas of symmetry breaking.  One 
of the great campaigns of particle physics over the next decade will 
be---must be---the quest for a complete understanding of electroweak 
symmetry breaking through an exploration of the 1-TeV scale.  The 
second theme is \textit{unity.} By unity I understand, of course, the 
hope that we can unify quarks and leptons and thus achieve a 
comprehensive theory of the strong, weak, and electromagnetic 
interactions with its consequent coupling constant unification.  But I 
also include the grander goals of unifying constituents and force 
particles, incorporating gravity into our theories of fundamental 
interactions, and reconciling quantum theory and relativity.  The 
third theme I call \textit{identity,} which incorporates the mystery 
of fermion masses and mixings, the origin and understanding of 
\textsf{CP} violation that we hope to gain through studies of the $K$ 
and $B$ systems, and the phenomenology of neutrino oscillations.  The 
problem of identity is as simple to state as this: what makes an 
electron an electron and a top quark a top quark?  

Our current understanding of flavor and family---of identity---is not 
so well developed as the visions we have for symmetry and unity.  I 
believe that the question of identity is an essential part of the 
physics that will determine the machine beyond the Large Hadron Collider.  
As we master, or at least gain a more mature understanding of, 
symmetry and unity, I expect that the questions of identity will 
increasingly define the agenda of particle physics.  Neutrino physics 
is at the center of those questions and
has an indispensable role to play in guiding our future.

\section{Ten Timely Questions in the Physics of Neutrino Oscillations}
We need answers to many questions in order to unravel the puzzle of 
neutrino physics.  The first question, to which the presumed answer 
motivates much of the current interest in neutrino physics, is \ldots

\subsection{Do neutrinos oscillate?}  
Many experiments have now used natural sources of neutrinos, neutrino 
radiation from fission reactors, and neutrino beams generated in 
particle accelerators to look for evidence of neutrino oscillation.  
The positive indications for neutrino oscillations fall into three 
classes \cite{janetc}:
\begin{enumerate}
	\item  Five solar-neutrino experiments report deficits with respect 
	to the predictions of the standard solar model: Kamiokande and 
	Super-Kamiokande using water-Cherenkov techniques, SAGE and GALLEX 
	using chemical recovery of germanium produced in neutrino 
	interactions with gallium, and Homestake using radiochemical 
	separation of argon produced in neutrino interactions with 
	chlorine.  These results suggest the oscillation $\nu_{e} 
	\rightarrow \nu_{x}$.

	\item  Five atmospheric-neutrino experiments report anomalies in the 
	arrival of muon neutrinos: Kamiokande, IMB, and Super-Kamiokande 
	using water-Cherenkov techniques, and Soudan II and MACRO using 
	sampling calorimetry.  The most striking result is the zenith-angle 
	dependence of the $\nu_{\mu}$ rate reported last year by Super-K 
	\cite{SKatm,SKLyon}.  These results suggest the oscillation 
	$\nu_{\mu} \rightarrow \nu_{\tau}\hbox{ or }\nu_{s}$.

	\item  The LSND experiment \cite{LSND} reports the observation of 
	$\bar{\nu}_{e}$-like events is what should be an essentially pure 
	$\bar{\nu}_{\mu}$ beam produced at the Los Alamos Meson Physics 
	Facility, suggesting the oscillation $\bar{\nu}_{\mu} \rightarrow
	\bar{\nu}_{e}$.  This result has not yet been reproduced by any 
	other experiment.
\end{enumerate}
A host of experiments have failed to turn up evidence for neutrino 
oscillations in the regimes of their sensitivity.  These results limit 
neutrino mass-squared differences and mixing angles.  In more than a 
few cases, positive and negative claims are in conflict, or at least 
face off against each other.  Over the next five years, many 
experiments will seek to verify, further quantify, and extend these 
claims.

Explanations other than neutrino oscillations have been advanced for 
some of these phenomena.  These include flavor-changing interactions 
\cite{fcint}, neutrino decay \cite{vbnudk}, violations of special 
relativity \cite{specrel}, and reservations about the standard solar 
model.  However, the most graceful interpretation of the oscillation 
evidence is that neutrinos have mass and neutrino flavors mix.

If neutrinos do oscillate, \ldots

\subsection{What are the neutrino masses?}

No one has ever weighed a neutrino.  The best kinematical 
determinations we have set upper bounds \cite{pdg} on the dominant 
neutrino species emitted in nuclear beta decay ($m_{\nu_{e}} \ltap 
15\evcc$), $\pi^{\pm}$ decay ($m_{\nu_{\mu}} < 0.19\mevcc$ at 
90\% CL), and $\tau$ decay ($m_{\nu_{\tau}} < 18.2\mevcc$ at 95\% 
CL).  Although there are prospects for improving these bounds 
\cite{alvaro}---and the measurement of a nonzero mass would 
constitute a real discovery---they are sufficiently large that it is 
of interest to consider indirect (nonkinematic) constraints from other 
quarters.

If neutrino lifetimes are greater than the age of the Universe, the 
requirement that neutrino relics from the Big Bang not overclose the 
Universe leads to a constraint on the sum of neutrino masses.  For 
relatively light neutrinos ($m_{\nu} \ltap\hbox{a few\mevcc}$), the 
total mass in neutrinos,
\begin{equation}
	m_{\mathrm{tot}} = \sum_{i}\half g_{i} m_{\nu_{i}} \; ,
	\label{totmass}
\end{equation}
where $g_{i}$ is the number of spin degrees of freedom of $\nu_{i}$ 
plus $\bar{\nu}_{i}$, sets the scale of the neutrino contribution to 
the mass density of the Universe, $\varrho_{\nu} = m_{\mathrm{tot}}n_{\nu} 
\approx 112m_{\mathrm{tot}}\cm^{-3}$.  If we measure 
$\varrho_{\nu}$ as a fraction of the critical density to close the 
Universe, $\varrho_{c} = 1.05 \times 10^{4}h^{2}\evcc \cm^{-3}$, 
where $h$ is the reduced Hubble parameter, then
\begin{equation}
	\Omega_{\nu} \equiv \frac{\varrho_{\nu}}{\varrho_{c}} = 
	\frac{m_{\mathrm{tot}}}{94h^{2}\evcc} \; .
	\label{omgnu}
\end{equation}
An assumed bound on $\Omega_{\nu}h^{2}$ then implies a bound on 
$m_{\mathrm{tot}}$.  A very conservative bound results from the 
assumption that $\Omega_{\nu}h^{2} < 1$: it is that $m_{\mathrm{tot}} < 
94\evcc$.

Recent observations \cite{costri} suggest that the total matter 
density is considerably smaller than the critical density, so that 
$\Omega_{m} \approx 0.3$.  If we fix $\Omega_{\nu} < \Omega_{m}$ and 
choose the plausible value $h^{2} = 0.5 \pm 0.15$, then we arrive at 
the still generous upper bound $m_{\mathrm{tot}} \ltap 19\evcc$.  
Taking into account the best (and model-dependent) information about 
the hot- and cold-dark-matter cocktail \cite{mst}, it seems likely that 
cosmology limits $m_{\mathrm{tot}} \ltap \hbox{a few}\evcc$.  It is 
worth remarking that the cosmological desire for hot dark matter has 
been on the wane.

If neutrinos do have mass, \ldots

\subsection{Is neutrino mass a sign of physics beyond the standard model?}
Until we have additional evidence, I believe that the right answer to 
this question is, ``It depends.''  All fermion masses and mixings are 
mysterious---by which I mean not calculable---within the standard 
model.  The gauge-boson masses are predicted in terms of the weak 
mixing parameter $\sin^{2}\theta_{W}$:
\begin{eqnarray}
	M_{W}^{2} & = & \frac{g^{2}v^{2}}{2} = 
	\frac{\pi\alpha}{G_{F}\sqrt{2}\sin^{2}\theta_{W}}
	\label{gbms}  \\
	M_{Z}^{2} & = & \frac{M_{W}^{2}}{\cos^{2}\theta_{W}} \; ,
	\nonumber
\end{eqnarray}
where $v = (G_{F}\sqrt{2})^{-1/2} = 246\gev$ sets the electroweak 
scale.  On the other hand, each fermion mass involves a new, unknown, 
Yukawa coupling.  For example, the term in the electroweak Lagrangian 
that gives rise to the electron mass is
\begin{equation}
	\lag_{\mathrm{Yuk}} = -\zeta_{e}\left[\bar{\mathsf{R}}(\varphi^{\dagger}\mathsf{L}) + 
	(\bar{\mathsf{L}}\varphi)\mathsf{R}\right]\; ,
	\label{eyuk}
\end{equation}
where $\varphi$ is the (complex) Higgs field and the left-handed and 
right-handed fermions are specified as
\begin{equation}
	\mathsf{L} = \left(
	\begin{array}{c}
		\nu_{e}  \\
		e
	\end{array}
	\right)_{\mathrm{L}}\; , \quad \mathsf{R} = e_{\mathrm{R}}
	\label{leptons}
\end{equation}
When the electroweak symmetry is spontaneously broken, the electron 
mass emerges as
\begin{equation}
	m_{e} = \zeta_{e}v/\sqrt{2}\; .
	\label{emass}
\end{equation}
The Yukawa couplings that reproduce the observed quark and lepton 
masses range over many orders of magnitude, from $\zeta_{e} \approx 
3 \times 10^{-6}$ for the electron to $\zeta_{t} \approx 1$ for the 
top quark.  Their origin is unknown.

In one sense, therefore, \textit{all fermion masses involve physics beyond 
the standard model.}  If we find that the electron neutrino has a 
Dirac mass reproduced by a Yukawa coupling $\zeta_{\nu_{e}} \approx 
10^{-10}$, perhaps nothing fundamentally new would be 
involved---though the mystery of fermion masses would still be a 
mystery.  We would still want to explain why neutrino masses are so 
small compared with charged-fermion masses.

It is worth remarking on another manifestation of the logical 
separation between the origin of gauge-boson masses and the origin of 
fermion masses.  The observation that a fermion mass is different from 
zero ($m_{f} \neq 0$) implies that the electroweak gauge symmetry 
\ewgg\ is broken, but electroweak symmetry breaking is only a 
necessary, not a sufficient, condition for the generation of fermion 
mass.  The separation is complete in simple technicolor \cite{TC}, the 
theory of dynamical symmetry breaking modeled on the 
Bardeen--Cooper--Schrieffer theory of the superconducting phase 
transition.

When we try to make sense of the Yukawa couplings $\zeta_{i}$, it is 
useful---probably essential---to keep in mind that according to 
unified theories, the pattern of fermion masses simplifies on high 
scales.  A theory based on $SU(5)$ with a specific 
symmetry-breaking pattern leads to simple quark-lepton mass relations 
at the unification scale, while an embedding of $SU(5)$ in $SO(10)$ 
reconciles a large $\nu_{\mu}$-$\nu_{\tau}$ mixing with the small 
quark mixing \cite{btauU}.

\subsection{Does the evidence require more than three neutrino species?}
Measurements of the invisible decay rate of the $Z$-boson, 
$\Gamma(Z^{0} \rightarrow \hbox{invisible})$, tell us with 
considerable precision that there are three light neutrinos with 
normal weak interactions: $N_{\nu} = 2.994 \pm 0.011$ \cite{pdg}.  The 
restriction to three light neutrinos does not preclude the existence of 
a ``sterile'' neutrino, $\nu_{s}$, that couples very feebly (so that 
$\Gamma(Z^{0} \rightarrow \nu_{s}\bar{\nu}_{s}) \ll \Gamma(Z^{0} 
\rightarrow \nu_{e}\bar{\nu}_{e}))$ or not at all (so that $Z^{0} 
\slashii{\rightarrow} \nu_{s}\bar{\nu}_{s}$) to the $Z^{0}$.  With a 
little stretching of error bars, a three-neutrino scenario can account 
for all the oscillation signals \cite{only3}, but \textit{at least 
four neutrinos} seem required to fit the central values quoted by the 
experiments \cite{fits}.

A simple argument indicates the necessity for more than three neutrino 
flavors.  If there are three mass eigenstates ($\nu_{1}, \nu_{2}, 
\nu_{3}$), then the sum of the squares of the mass differences, 
suitably defined, must vanish:
\begin{equation}
	\sum \delta M_{ij}^{2} = (M_{3}^{2} - M_{2}^{2}) +
	(M_{2}^{2} - M_{1}^{2}) + (M_{1}^{2} - M_{3}^{2}) = 0.
	\label{zipident}
\end{equation}
However, experiments that study solar neutrinos, atmospheric 
neutrinos, and accelerator-generated muon antineutrinos seem to 
require three very different values of $\delta M^{2}$:
\begin{eqnarray}
	|\delta M^{2}|_{\mathrm{solar}} & \approx & 10^{-10}\ev^{2}\hbox{ 
	or }10^{-5}\ev^{2}\; ;
	\nonumber  \\
	|\delta M^{2}|_{\mathrm{atm}} & \approx & 10^{-3}\hbox{ - } 
	10^{-2}\ev^{2}\; ;
	\label{dm2s}  \\
	|\delta M^{2}|_{\mathrm{LSND}} & \approx & 10^{-1}\hbox{ - 
	}10^{1}\ev^{2}\; .
	\nonumber
\end{eqnarray}
No choice of signs allows us to sum these three scales of $\delta 
M^{2}$ to zero.  For the moment, the conclusion that there must be 
more than three neutrino species is not a robust result, because the 
LSND anomaly has not (yet!) been confirmed by an independent 
experiment, and the determination of the preferred range of $|\delta 
M^{2}|$ has an impressionistic quality in all experiments 
\cite{gaufre}.  It is, however, a conclusion lingering on the horizon 
that we cannot entirely ignore.  Many theorists, motivated by their 
convictions about mass patterns, or their doubts about the LSND 
experiment, or their fear of opening Pandora's box, choose to put 
aside for the moment the conclusion that we require at least four 
neutrino species.  It is fine to wait and see, but we must also be 
ready to take the evidence as it comes.

If the evidence from mass differences remains inconclusive, \ldots

\subsection{Can we find evidence for (or against) a sterile neutrino?}
If a neutrino oscillates, it is essential that we learn what it 
oscillates into.  For the moment, it is common practice to suppose 
that each oscillation effect is governed---in first approximation---by 
a single transition.  The best way of confirming an oscillation 
between two $SU(2)_{L}$ flavors is by observing the 
\textit{appearance} of a species not in the initial beam.  Until now, 
the only appearance experiment we have is the unconfirmed LSND 
observation.  In view of the suspicion \cite{steratm} that the 
atmospheric neutrino anomaly reflects a $\nu_{\mu} \rightarrow 
\nu_{\tau}$ transition, it is very important to carry out 
long-baseline experiments capable of observing $\tau$ appearance.  A 
comparison of the neutral-current / charged-current ratio---or a 
measurement of an exclusive neutral-current rate---at a far 
detector gives information about the total flux of $SU(2)_{L}$ 
flavors.  It can be a valuable tool to discriminate between 
flavor-flavor oscillations and flavor-sterile oscillations, and is the 
goal of many experiments in the next round.

If there is one sterile neutrino---which must be light enough to mix 
with the $SU(2)_{L}$ neutrinos---why shouldn't there be (at least) 
three?  What do we need to know about a sterile neutrino?  What sort 
of experiments can tell us?

\subsection{Could neutrino masses be special?}
Alone among the known fermions, the neutral neutrino can be its own 
antiparticle.  This fact opens the possibility of several varieties 
of neutrino masses.  Let us begin by making a chiral decomposition of 
the neutrino's Dirac spinor,
\begin{equation}
	\psi = \half(1-\gamma_{5})\psi + \half(1 + \gamma_{5})\psi \equiv 
	\psi_{\mathrm{L}} + \psi_{\mathrm{R}}\; ,
	\label{chidcom}
\end{equation}
and remarking that the charge conjugate of a right-handed field is 
left-handed:
\begin{equation}
	\psi_{\mathrm{L}}^{c} \equiv (\psi^{c})_{\mathrm{L}} =
	(\psi_{\mathrm{R}})^{c}\; .
	\label{rtol}
\end{equation}
What are the possible forms that neutrino mass terms might take?

A Dirac mass term connects the left-handed and right-handed components 
of the same field.  It is represented by the Lagrangian term
\begin{equation}
	\lag_{D} = D(\bar{\psi}_{\mathrm{L}}\psi_{\mathrm{R}}+ 
	\bar{\psi}_{\mathrm{R}}\psi_{\mathrm{L}}) = D\bar{\psi}\psi\; ,
	\label{Dlag}
\end{equation}
which implies a mass eigenstate $\psi = \psi_{\mathrm{L}} + 
\psi_{\mathrm{R}}$.  The Dirac mass eigenstate is invariant under the 
global phase rotation $\nu \rightarrow e^{i\theta}\nu$, $\ell 
\rightarrow e^{i\theta}\ell$, so that lepton number is conserved.

Majorana mass terms connect the left-handed and right-handed 
components of conjugate fields.  They are represented by the 
Lagrangian terms
\begin{eqnarray}
	-\lag_{MA} & = & A(\bar{\psi}_{\mathrm{R}}^{c}\psi_{\mathrm{L}} + 
	\bar{\psi}_{\mathrm{L}}\psi_{\mathrm{R}}^{c}) = A \bar{\chi}\chi
	\nonumber  \\
	-\lag_{MB} & = & 
	B(\bar{\psi}^{c}_{\mathrm{L}}\psi_{\mathrm{R}} + 
	\bar{\psi}_{\mathrm{R}}\psi^{c}_{\mathrm{L}}) = B\bar{\omega}\omega\; ,
	\label{Majlag}
\end{eqnarray}
for which the mass eigenstates are
\begin{eqnarray}
	\chi & \equiv & \psi_{\mathrm{L}} + \psi^{c}_{\mathrm{R}} = \chi^{c}
	= \psi_{\mathrm{L}} + (\psi_{\mathrm{L}})^{c} \; ,
	\nonumber  \\
	\omega & \equiv & \psi_{\mathrm{R}} + \psi^{c}_{\mathrm{L}} = \omega^{c}
	= \psi_{\mathrm{R}} + (\psi_{\mathrm{R}})^{c} \; .
	\label{Majeigen}
\end{eqnarray}
The coupling of conjugate fields in the Majorana mass terms violates 
lepton number by two units.  Accordingly, Majorana neutrinos can 
mediate neutrinoless double-beta decays ($\beta\beta_{0\nu}$) in 
heavy nuclei,
\begin{equation}
     (Z,A) \rightarrow (Z+2,A) + e^{-} + e^{-} \; .
\end{equation}
Detection of neutrinoless double-beta decay would offer decisive 
evidence for the Majorana nature of the neutrinos \cite{doublebetarev}. 

The Heidelberg--Moscow experiment has recently set the most stringent 
limit on a \textit{Majorana} neutrino mass \cite{doublebeta}.  Their 
lower limit on the half-life for neutrinoless double-beta decay of 
$^{76}$Ge, $t_{1/2}^{\beta\beta_{0\nu}} \ge 5.7 \times 
10^{25}\hbox{ yr}$ at 90\% CL, restricts an effective Majorana 
neutrino mass to be $\ltap 0.2\evcc$.

With both Dirac and Majorana mass terms, the neutrino mass 
contribution to the Lagrangian is
\begin{eqnarray}
     -\lag_{DM} & = & \half D(\bar{\chi}\omega + 
     \bar{\omega}\chi) + A\bar{\chi}\chi + B\bar{\omega}\omega 
     \nonumber \\
      & = & (\bar{\chi}, \bar{\omega}) \left( 
      \begin{array}{cc}
      	A & \half D  \\
      	\half D & B
      \end{array}
      \right) \left( 
      \begin{array}{c}
      	\chi  \\
      	\omega
      \end{array}
      \right)\label{DMlag}\; ,
\end{eqnarray}
which has mass eigenvalues
\begin{equation}
M_{2,1} = \frac{A + B \pm \sqrt{(A - B)^{2} + D^{2}}}{2} \; ,
\label{DMevals}
\end{equation}
which is to say two Majorana mass eigenstates.

A favorite realization of the Dirac--Majorana mass alternative is the 
so-called see-saw mechanism \cite{bascule}, which offers the prospect of a connection 
to high-scale physics, and thus an opening to true physics beyond the 
standard model.  Let us assume that the Dirac mass $D$ takes a 
typical value of a lepton mass in the electroweak theory.  If the 
left Majorana mass $A \approx 0$ is negligible, and the right 
Majorana mass $B \gg |D|$, then the two mass eigenvalues are given by
\begin{equation}
M_{2,1} = \frac{B \pm \sqrt{B^{2} + D^{2}}}{2} \; .
\label{eq:2masses}\end{equation}
To excellent approximation, we have
\begin{equation}
M_{2} = B, \quad M_{1} = -\frac{D^{2}}{4B}\; .
\label{seesawevals}
\end{equation}
Because $M_{1}$ is small compared with the typical lepton mass $D$, 
this scheme offers a ``natural'' explanation for the observed strong 
inequality, $m_{\nu} \ll m_{e}$.  It also leads naturally to a 
sterile neutrino; but notice that the sterile neutrino is heavy, not 
the light sterile neutrino needed to accommodate all the hints of 
neutrino oscillation.

\subsection{How could light sterile neutrinos arise?}
If the data do lead us to consider mixing between the $SU(2)_{L}$ 
neutrinos and a sterile neutrino, what are the mechanisms that might 
produce a light sterile neutrino, and what other implications would 
they have for neutrino physics and beyond.  Among sources of light 
scalar neutrinos that have been investigated, are the radiative 
generation of $m_{\nu}$ and induced masses that arise through 
$R$-parity violation in supersymmetry.  If neutrino masses are 
generated through loop diagrams, neutrinoless double-beta decay does 
not arise in general.  $R$-parity--violating supersymmetry has a great 
number of potentially observable consequences.  It would be useful to 
focus on these---and other---mechanisms for light scalar neutrinos, 
to understand what demands they would put on a neutrino factory's 
performance.

\subsection{Are neutrino mixing angles large?  maximal?}
Although two of the favored interpretations for the solar-neutrino 
deficit feature large mixing angles, the possibility of resonant 
conversion within the varying matter profile of the Sun allows a 
small-mixing-angle solution.  The density profile of the Earth does 
not naturally allow resonant conversion of atmospheric neutrinos over 
a broad range of energies, and so it is generally accepted that large 
mixing is required to account for the atmospheric neutrino anomaly.  
This conclusion and the existence of the large-angle solar solutions 
discourage the formerly traditional belief that the structure of the 
neutrino mixing matrix should be similar to that of the familiar quark 
mixing matrix.

If the mixing among neutrino flavors that accounts for the atmospheric 
neutrino anomaly is large---or, indeed, maximal---should we interpret 
it as large flavor-sterile mixing, or as large flavor-flavor mixing?
In either case, it seems natural to take the large mixing as an 
important clue to neutrino properties \cite{celw}.

\subsection{Do neutrino masses probe large extra dimensions?}
It is a longstanding dream of string theory that string modes in the 
dimensions beyond the known $3+1$ spacetime dimensions might 
determine the properties of the quarks and leptons.  On this 
interpretation, the structure of the Calabi--Yau manifolds in the 
small dimensions is reflected in the spectrum of what we take, at our 
limited resolution, to be elementary particles.  It offers a novel 
approach to the problem of identity.

Over the past eighteen months, the apparently preposterous idea that 
some of the extra spatial dimensions might be perceivably large has 
shown itself to be not at all easy to rule out, and both entertaining 
and informative to consider.  A number of authors have suggested that 
the physics of extra dimensions might give rise to neutrino properties 
and oscillations \cite{ddg}.  How can we test these 
mechanisms?

\subsection{Can we detect \textsf{CP} violation in neutrino mixing?}
Using the beams available at a neutrino factory, it will be of great 
interest to compare
\begin{eqnarray}
 \nu_{e} \rightarrow \nu_{\mu} & \mathit{~vs.~} & \nu_{\mu} \rightarrow 
 \nu_{e} \; , \nonumber \\
 \nu_{e} \rightarrow \nu_{\mu} & \mathit{vs.} & 
 \bar{\nu}_{\mu} \to \bar{\nu}_{e} \; , \\ 
 \nu_{e} \rightarrow \nu_{\mu} & \mathit{vs.} & \bar{\nu}_{e} \to 
 \bar{\nu}_{\mu} \; .\nonumber
\end{eqnarray}
Although oscillation probabilities are insensitive to the signs of 
$\delta M^{2}$, matter effects do depend on the ordering of neutrino 
masses, so the observation of Earth matter effects could help us 
determine the pattern of neutrino masses completely.  Matter 
effects can mimic some of the unequal rates induced by \textsf{CP} 
violation, so it is essential to understand them for engineering 
purposes.

In representative scenarios for the pattern of neutrino masses, what 
constitutes a definitive program of measurements to separate matter 
effects from \textsf{CP} violation?  What implications does that 
program have for the capabilities of detectors and for muon energy and 
the ability to manipulate muon polarization?  The discovery of 
\textsf{CP} violation in the neutrino system would have profound 
implications for questions of identity, and is worth pursuing 
aggressively, if the pattern of mixing makes it a plausible target.  
Introductions to the problem can be found in References 
\cite{bgw,pilar}.

\section{Concluding Remarks}
I commend the workshop organizers and our Lyonnais hosts for putting 
together a stimulating and enjoyable program.  I particularly want to 
thank Serguey Petcov and Belen Gavela for their indispensable 
contributions to the theory working group.  I am grateful to the CERN 
Theoretical Studies Division for warm hospitality following $\nu$Fact 
'99.  Fermilab is operated by Universities Research Association Inc.  
under Contract No.  DE-AC02-76CH03000 with the United States 
Department of Energy.

Now, let us get down to work to set out our physics goals and 
understand what a neutrino factory should be.  As we do that, I would 
like to return to one of my opening questions: Is a neutrino factory 
the only way---or the best way---to provide information we so urgently 
need about neutrinos, flavor, and identity?


\begin{thebibliography}{999}
\bibitem{gaufre} In particular, let us note that the surprising 
observation by the Homestake experiment of a smaller solar-neutrino 
interaction rate than seen by the water-Cherenkov detectors, despite 
its lower-energy threshold, has a special influence on the allowed 
solutions.  Time will tell whether this is a crucial fact or not.
\bibitem{steratm} At this workshop, we have seen a gentle hint (at the 
2-$\sigma$ level) from the Super-Kamiokande experiment that their 
atmospheric neutrino anomaly prefers oscillation into $\nu_{\tau}$ 
over oscillation into $\nu_{s}$ [see Ref.\ \cite{SKLyon}].  The 
analysis is based on the different survival probability of 
$\nu_{\tau}$ and $\nu_{s}$ propagating through matter, as suggested by 
P.  Lipari and M.  Lusignoli, \pr{58}{073005}{98} \hepph{9803440}.  
For similar ideas, see G.  Barenboim and F.  Scheck, 
\pl{B450}{189}{99} \hepph{9812351}.
\end{thebibliography}
\end{document}